\documentclass[10pt,a4paper,twocolumn,aps,pra,showpacs,superscriptaddress]{revtex4-1}
\pdfoutput=1
\usepackage[utf8x]{inputenc}
\usepackage{ucs}
\usepackage{amsmath}
\usepackage{amsfonts}
\usepackage{amssymb}
\usepackage{makeidx}
\usepackage{cellspace,booktabs}
\usepackage{natbib}
\usepackage{lipsum}
\usepackage{hyperref}

\usepackage{color}
\definecolor{light-gray}{gray}{0.55}

\usepackage{microtype}
\usepackage{bm}

\usepackage{graphicx}

\renewcommand{\dag}{^{\dagger}}
\newcommand{\exv}[1]{ \langle #1 \rangle }

\newcommand{\bra}[1]{ \langle #1 \rvert }
\newcommand{\ket}[1]{ \lvert #1 \rangle}
\newcommand{\braket}[2]{\langle #1 \vert #2 \rangle }

\newcommand{\pfrac}[2]{\frac{\partial #1}{\partial #2}}

\newcommand{\wa}{\omega_a}
\newcommand{\wc}{\omega_r}

\newcommand{\phivec}{v}

\begin{document}

\begin{abstract}
The interaction of light and matter is often described by the exchange of single excitations. When the coupling strength is a significant fraction of the system frequencies, the number of excitations are no longer preserved and that simple picture breaks down. This regime is known as the ultrastrong coupling regime and is characterized by non-trivial light-matter eigenstates and complex dynamics. In this work, we propose to use a an array Josephson junctions to increase the impedance of the light mode enabling ultrastrong coupling to a transmon qubit. We show that the resulting dynamics can be generated and probed by taking advantage of the multi-mode structure of the junction array. This proposal relies on the frequency tunability of the transmon and, crucially, on the use of a low frequency mode of the array, which allows for non-adiabatic changes of the ground state. 
\end{abstract}

\date{\today}

\author{Christian Kraglund Andersen}
\thanks{Email: ctc@phys.au.dk}
\affiliation{Department of Physics and Astronomy, Aarhus University, DK-8000 Aarhus C, Denmark}

\author{Alexandre Blais}
\affiliation{Institut Quantique, Université de Sherbrooke, Sherbrooke, Québec J1K 2R1, Canada}
\affiliation{Canadian Institute for Advanced Research, Toronto, Ontario M5G 1Z8, Canada}


\title{Ultrastrong coupling dynamics with a transmon qubit}

\maketitle

\section{Introduction}

Cavity quantum electrodynamics allows for the study of light-matter interaction at the level of single atoms interacting with a single photon, both confined in a high-quality cavity. In practice, this interaction is typically due to the coupling of the light's electric field to the electric dipole moment of the atom~\cite{haroche2006exploring}. When only a single mode of light and only two atomic levels are relevant, this situation can be described by the Jaynes-Cummings Hamiltonian ($\hbar = 1$),
\begin{align}
H_{\mathrm{JC}} =  \wc \, a\dag a + \frac{\wa}{2} \sigma_z + g (a \dag \sigma_- + a \sigma_+), \label{eq:simplejc}
\end{align}
where $\wc$ is the cavity frequency, $\wa$ the atomic frequency and $g$ the electric-dipole coupling. In this expression, $a$ ($a\dag$) is the photon annihilation (creation) operator and $\sigma_i$ are the Pauli matrices for the atomic levels. The Jaynes-Cummings Hamiltonian describes the exchange of a single quanta between the field and the atom leading to Rabi oscillations with angular frequency $2g$. The strong coupling regime is achieved when the coupling, $g$, is much larger than the dissipation rates of the system. This Jaynes-Cummings Hamiltonian can be realized in a wide variety of physical systems such as Rydberg atoms~\cite{haroche2006exploring}, quantum dots~\cite{imamog1999quantum}, trapped ions~\cite{RevModPhys.75.281, PhysRevA.56.2352} and superconducting circuits~\cite{wallraff2004strong}.

The Jaynes-Cummings Hamiltonian is, however, only an approximation of the Rabi Hamiltonian describing coupling between the cavity electric field, $E_0(a\dag+a)$, and the atomic dipole moment, $d_0\sigma_x$,
\begin{align}\label{eq:simplerabi}
H_{\mathrm{Rabi}} = \wc \, a\dag a + \frac{\wa}{2} \sigma_z + g (a \dag + a) \sigma_x, 
\end{align}
where $g=d_0E_0$. The Jaynes-Cummings Hamiltonian is a good approximation to $H_{\mathrm{Rabi}}$ when the coupling, $g$, is smaller than the system frequencies, $g\ll \wa, \wc$. In this situation, the fast rotating term $a\dag\sigma_++a\sigma_-$ appearing in the Rabi Hamiltonian can safely be dropped using the rotating wave approximation (RWA) and we recover Eq.~\eqref{eq:simplejc}. While more challenging to realize, there has recently been much attention to the situation where this approximation is no longer valid. This so-called ultrastrong regime, realized when the coupling strength approaches the system frequencies, differs remarkably from the Jaynes-Cummings regime~\cite{ciuti:2006a, PhysRevLett.107.100401, PhysRevA.82.062320, PhysRevLett.105.023601, PhysRevLett.105.263603, cao2011qubit, PhysRevA.92.053823, gunter2009sub}. Most significantly, while the ground state of  $H_{\mathrm{JC}}$ is simply the product of the atomic ground state and vacuum of the field, the ground state of the Rabi Hamiltonian is an entangled atom-field state with a non-zero average photon number. From a practical point of view, this regime could also be useful in the context of quantum information processing~\cite{PhysRevLett.113.263604, PhysRevLett.108.120501, PhysRevB.91.064503, felicetti2015parity}.

Superconducting quantum circuits form a promising platform to realize and study this novel light-matter coupling regime. In particular, realization of the ultrastrong coupling regime with flux qubits, acting as the atom, coupled to a microwave resonator have been theoretically studied~\cite{PhysRevA.80.032109} and experimentally implemented~\cite{niemczyk2010circuit, forn2010observation, forn-diaz:2016a, yoshihara:2016a}. These experiments have primarily probed the spectral properties of the ultrastrong regime. A next step is to probe the dynamics of the system in this regime and, moreover, to probe its non-trivial ground state. With the system starting in its ground state, an approach is to non-adiabatically tune the coupling strength $g$ from the ultrastrong coupling regime to the strong coupling regime. The system will readjust to this change by emitting photons as it relaxes back to its new ground state. Observing these photons would constitute a clear signature of the non-trivial nature of the ultrastrong coupling ground state. With system frequencies around 10 GHz~\cite{PhysRevA.80.032109,niemczyk2010circuit, forn2010observation,yoshihara:2016a}, this however requires changes in system parameters of the order of 10 pico-seconds. In practice, this therefore appears to be extremely challenging. 

In this work, we address this problem by working with a low-frequency mode of a microwave cavity. We focus on the transmon qubit~\cite{PhysRevA.76.042319} capacitively coupled to an array of Josephson junctions realizing an inductance with a dissipationless impedance larger than the resistance quantum~\cite{PhysRevLett.109.137002, PhysRevLett.109.137003, PhysRevB.92.104508}. Together with its capacitance to ground, this superinductance plays the role of a multi-mode cavity.  With $g/\wc \propto \sqrt{Z_0}$, where $Z_0$ is the cavity impedance~\cite{devoret:2007a,PhysRevA.80.032109}, this approach allows for large qubit-mode coupling strengths. Moreover, by using a low-frequency mode of the array, it is possible to realize ultrastrong coupling with only a moderately large coupling strength. In this situation, fast changes of system parameters are possible and allow for the observation of signatures of the ultrastrong coupling in the dynamics of the combined system. This dynamic can then be probed by taking advantage of the presence of multiple modes of the array and their cross-Kerr interaction~\cite{PhysRevLett.109.137002}.

The paper is organized as follows:  We derive in Sec.~\ref{sec:lagran} the Hamiltonian of the system, taking into account the multi-mode structure of the array. In Sec.~\ref{sec:parameters}, we identify parameters to reach the ultrastrong coupling regime. In Sec.~\ref{sec:dynamics} the dynamics of the ultrastrong coupling regime is investigated. Finally, Sec.~\ref{sec:conc} concludes the paper.
\begin{figure}[t]
\includegraphics[width=0.8\linewidth]{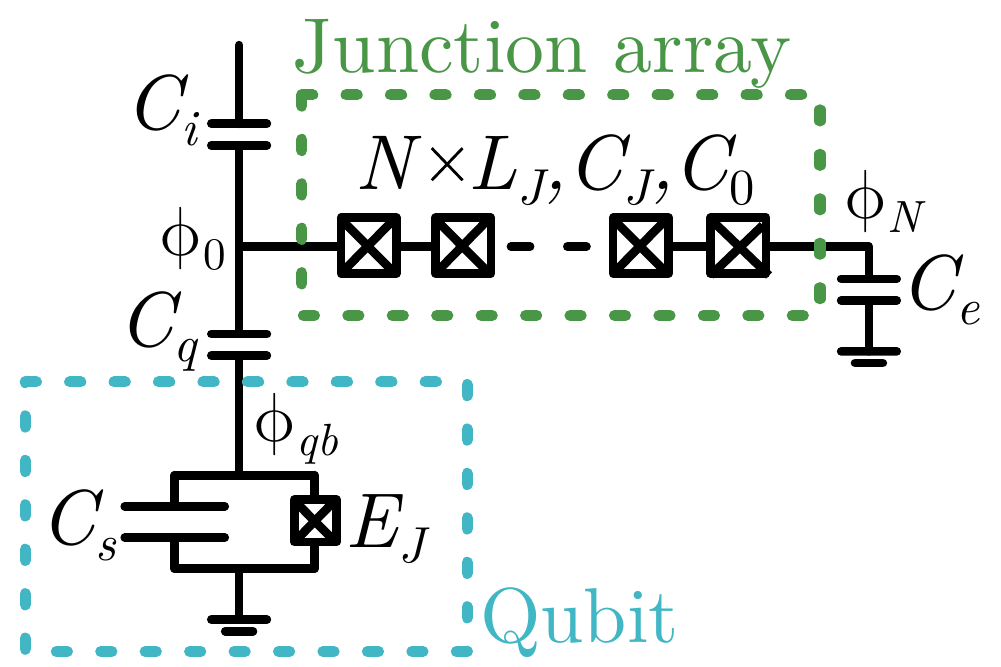}
\caption{Circuit representation of the system. An array of $N$ Josphson junctions treated as a series of inductors with  inductance $L_J$ and capacitance $C_J$. These junctions also have a parasitic capacitance to ground, $C_0$.  The array is interaction via $C_q$ with a transmon qubit characterized by the capacitance $C_s$ and the Josephson energy $E_J$. The flux node at the transmon is denoted $\phi_{qb}$ and the nodes of the array goes from $\phi_0$ to $\phi_N$.
} \label{fig:lumped}
\end{figure}

\section{Transmon coupled to a Josephson junction array}
\label{sec:lagran}

We consider the circuit of Fig.~\ref{fig:lumped} which consists of a transmon qubit~\cite{PhysRevA.76.042319} coupled to an array of $N$ Josephson junctions~\cite{PhysRevLett.109.137002}. The transmon qubit is characterized by the Josephson energy $E_J$ and the capacitance $C_s$, which for simplicity we take to include both the shunt capacitance and the junction capacitance. We assume the junctions forming the array to have a large Josephson energy such that, to a good approximation, they behave as weakly nonlinear inductances. These junctions are then characterized by their Josephson  inductance $L_J$ and junction capacitance $C_J$. Following Refs.~\cite{PhysRevA.86.013814, PhysRevLett.108.240502}, the nonlinearity of the array junctions will be perturbatively reintroduced at a later step. Moreover, we take into account the capacitance to ground $C_0$ of the islands formed between the array junctions. The capacitance $C_q$ couples the qubit to the array and will largely control their interaction strength. Finally, $C_i$ is a capacitance to an external control field which will be used to probe the system and $C_e$ is the capacitance to ground of the last array island, which can be constructed arbitrarily~\cite{PhysRevLett.109.137002}.

Following the standard approach~\cite{devoret1995quantum}, the circuit Lagrangian reads
\begin{align}
\mathcal{L} 
&= \sum_{n=0}^{N-1} \Big[\frac{C_J}{2} (\dot{\phi}_n {-} \dot{\phi}_{n+1})^2 - \frac{1}{2L_J} (\phi_n {-} \phi_{n+1})^2 + \frac{C_0}{2} \dot{\phi}_n^2 \Big]\nonumber\\
&\quad + \frac{C_i}{2} \dot{\phi}_0^2 + \frac{C_e}{2} \dot{\phi}_{N}^2 + \frac{C_q}{2}(\dot{\phi}_0 - \dot{\phi}_{qb})^2 \label{eq:philagran_pre}\\
&\quad + \frac{C_s}{2} \dot{\phi}_{qb}^2 + E_J \cos \big(\phi_{qb}\, / \varphi_0 \big) \nonumber\\
&= \dot{\vec{\phi}}^{\,T} \frac{\bm{C}}{2} \dot{\vec{\phi}} - \vec{\phi}^{\,T} \frac{\bm{L}}{2} \vec{\phi} + E_J \cos \big(\phi_{qb}\, / \varphi_0 \big), \label{eq:philagran}
\end{align}
where $\phi_n$ is the node flux of the $n$th island of the array, $\phi_{qb}$ the node flux of the transmon's island and $\varphi_0 = \Phi_0/2\pi$ with the magnetic flux quantum $\Phi_0 = h/(2e)$. In the last line of this equation, we have defined the vector \mbox{$\vec{\phi} = \{ \phi_0, \, \phi_1, \, \ldots \, , \phi_N, \phi_{qb}\}^{T}$} of length $N+2$, and the capacitance and inductance matrices $\bm{C}$ and $\bm{L}$ such that Eq.~\eqref{eq:philagran} reproduces Eq.~\eqref{eq:philagran_pre}.

Ignoring the non-linear term proportional to $E_J$ for the moment, this Lagrangian leads to the Euler-Lagrange equation
\begin{align}
\ddot{\vec{\phi}} = - \bm{C}^{-1} \bm{L} \vec{\phi} \equiv \bm{\Omega}^2 \vec{\phi},
\end{align}
which has the qubit mode $\vec{\phi}_{qb} = \{0,\, \ldots\, , 0,\phi_{qb}\}^T$ as one of the eigenvectors. Since the qubit's Josephson energy is not included in the matrix $\bm{\Omega}^2=  - \bm{C}^{-1} \bm{L}$ of size \mbox{$N{+}2{\times}N{+}2$}, this mode has zero eigenvalue and can easily be identified. A convenient basis to treat the array and the qubit separately is obtained by finding the eigenvectors of the \mbox{$N{+}1{\times}N{+}1$} block matrix of $\bm{\Omega}^2$ that does not relate to $\vec{\phi}_{qb}$.
We refer to these eigenvectors as $\vec{\phivec}_k$, such that the flux across the array is given as $\vec{\phi}(t) = \sum_k \phi_k(t) \vec{\phivec}_k$ with the time-dependence written explicitly. With this approach, the array modes are already renormalized by the qubit capacitances, $C_q$ and $C_s$. In the basis of these eigenmodes, the Lagrangian Eq.~\eqref{eq:philagran} takes the simple form 
\begin{equation}
\begin{split}
\mathcal{L} &= \sum_k\Big[ \frac{C_k}{2} \dot{\phi}_k^2 - \frac{1}{2L_k} \phi_k^2 - C_q \phivec_k(0) \, \dot{\phi}_k \dot{\phi}_{qb} \Big] \\
&\quad + \frac{C_q+C_s}{2} \dot{\phi}_{qb}^2 + E_J \cos \big(\phi_{qb}\, / \varphi_0 \big),
\end{split}
\end{equation}
where $\phivec_k(n)$ denotes the $n$th entry of the eigenvector $\vec{\phivec}_k$. In the above expression, we have defined the mode capacitance $C_k$ and mode inductance $L_k$ as
\begin{align}
C_k = \vec{\phivec}_k^{\,T} \bm{C} \vec{\phivec}_k, && L_k^{-1} = \vec{\phivec}_k^{\,T} \bm{L} \vec{\phivec}_k.
\end{align}
With these definitions, the eigenmode frequencies take the usual form $\omega_k = 1/\sqrt{L_k C_k}$. 

To obtain the associated Hamiltonian, we first identify the conjugate variables
\begin{align}
q_k &= \pfrac{\mathcal{L}}{\dot{\phi}_k} = C_k\dot{\phi}_k - C_q{\phivec}_k(0)\, \dot{\phi}_{qb}, \label{eq:qk} \\
q_{qb} &= (C_q + C_s)\dot{\phi}_{qb} -\sum_k C_q{\phivec}_k(0)\, \dot{\phi}_k. \label{eq:qqb}
\end{align}
Introducing $\tilde{q}$ and $\tilde{\phi}$ as the row vectors of entries $q_{k\,(qb)}$ and $\phi_{k\,(qb)}$, the above expressions can be written in compact vector form as
\begin{align}
\tilde{q} = \bm{\tilde{C}} \dot{\tilde{\phi}}. \label{eq:tildeC}
\end{align}
We also define $\bm{\tilde{L}}$, the diagonal matrix of matrix elements $1/L_k$. Using this notation a Legendre transformation is performed and the Hamiltonian reads
\begin{align}
H = \tilde{q}^{\,T} \frac{\bm{\tilde{C}}^{-1}}{2} \tilde{q} + \tilde{\phi}^{\,T} \frac{\bm{\tilde{L}}}{2} \tilde{\phi} - E_J \cos \phi_{qb}, \label{eq:hamil}
\end{align}
with the capacitances and inductances for mode $k$ given by the diagonal entries of the matrices,
\begin{align}
\tilde{C}_k^{-1} = \bm{\tilde{C}}_{[k,k]}^{-1}, && \tilde{L}_k^{-1} = \bm{\tilde{L}}_{[k,k]}.
\end{align}

The Hamiltonian of Eq.~\eqref{eq:hamil} can be expressed as the sum of a qubit Hamiltonian, $H_{qb}$, an array Hamiltonian, $H_{array}$, and their coupling, $H_c$. The qubit Hamiltonian takes the standard form
\begin{align}
H_{qb} = 4 E_C \hat{n}^2 - E_J \cos \big( {\phi}_{qb} / \varphi_0 \big), \label{eq:transmonqb}
\end{align}
where $-2e\hat{n} = \hat{q}_{qb}$ and $E_C = e^2 /(2\tilde{C}_{qb})$. In the transmon regime, $E_J/E_c \gg 1$, this Hamiltonian can be approximated as~\cite{PhysRevA.76.042319} 
\begin{align}\label{eq:Hqb1}
H_{qb} \approx \omega_a b\dag b - \frac{E_C}{2}b\dag b\dag bb, 
\end{align}
with the transmon frequency $\omega_{a} = \sqrt{8E_J E_C}-E_C$ and where we have introduced
\begin{align}
\phi_{qb} &= \varphi_0 \left(\frac{2 E_C}{E_J}\right)^{1/4}(b\dag+b),\\
\quad
n_{qb} &= i\left(\frac{E_J}{32 E_C}\right)^{1/4}(b\dag-b).
\end{align}
While the above form is useful in simplifying analytical expressions, all numerical calculations in this paper are based on the exact diagonalization of Eq.~\eqref{eq:transmonqb}.

Expressing the mode operators $q_k$ and $\phi_k$ in terms of the creation (annihilation) operators $a_k\dag$ ($a_k$) for mode $k$ of the array
\begin{align}
{q}_k = i\sqrt{\frac{  \tilde{\omega}_k \tilde{C}_k}{2}}  (a_k\dag - a_k),\\
{\phi}_k = \sqrt{\frac{ \tilde{\omega}_k \tilde{L}_k}{2}}  (a_k\dag + a_k), \label{eq:phihat}
\end{align}
the array Hamiltonian takes the standard from $H_{array} = \sum_k \tilde{\omega}_k a_k\dag a_k$. In this expression, the mode frequencies are $\tilde{\omega}_k = 1/(Z_k \tilde{C}_k)$ where $Z_k = \sqrt{\tilde{L}_k/\tilde{C}_k}$ is the characteristic impedance of mode $k$~\cite{PhysRevA.86.013814,devoret:2007a}. The frequencies $\tilde{\omega}_k$ differ slightly from $\omega_k$ due to the off-diagonal elements of $\bm{\tilde{C}}$. As can be seen from the first term of Eq.~\eqref{eq:hamil}, these terms also causes a small coupling between the array modes. This mode-mode coupling is due to the qubit-array interaction and is analogous to a multi-mode $A^2$-term~\cite{PhysRevLett.107.113602, nataf2010no}.  Omitting array junctions nonlinearities, the Hamiltonian then takes the form
\begin{align}
{H}
& = H_{qb} + H_{array} + \sum_k g_k (b\dag - b) (a_k\dag - a_k^{\phantom{\dagger}}) \nonumber \\
&\quad + \sum_{k\neq l}  G_{kl} (a_k\dag - a_k^{\phantom{\dagger}}) (a_l\dag - a_l^{\phantom{\dagger}}). \label{eq:Hhat}
\end{align}
In the transmon regime $E_J/E_C \gg 1$, the qubit-array coupling strength takes the form
\begin{align}
g_k = \Bigg(\frac{8E_J}{E_C} \Bigg)^{\frac{1}{4}} \sqrt{\frac{1}{2 Z_k}} e \;\bm{\tilde{C}}^{-1}_{[k,qb]}. \label{eq:g}
\end{align}
As expected, we find that $g_k/\tilde{\omega}_k \propto \sqrt{Z_k}$~\cite{devoret:2007a,PhysRevA.80.032109}.  As already mentioned, in addition to a qubit-array coupling, Eq.~\eqref{eq:Hhat} also contains a mode-mode interaction given by
\begin{align}
G_{kl} = \sqrt{\frac{1 }{2 Z_k}} \sqrt{\frac{1}{2 Z_l}} \;\bm{\tilde{C}}^{-1}_{[k,l]}/2.
\end{align}
In practice the frequency difference between modes is such that $\tilde{\omega}_l-\tilde{\omega}_k \gtrsim 100 G_{kl}$, evaluated using the parameters used in Sec.~\ref{sec:parameters}. Due to the small magnitude of these $G_{kl}$, we can neglect their renormalization of the mode-frequencies.

To finalize the derivation of the system Hamiltonian, we now include the array junction nonlinearities following the approach of Refs.~\cite{PhysRevA.86.013814, PhysRevLett.108.240502}. Taking advantage of the weak nonlinearity of these junctions, we consider only the fourth order expansion of the cosine potential of each array junction leading to the non-linear potential
\begin{align}
U_{nl} = - \frac{1}{24L_J \phi_0} \sum_{n=0}^{N-1} (\phi_n - \phi_{n+1})^4. \label{eq:Unl}
\end{align}
Expressing this in the eigenmode basis, using the mode creation and annihilation operators, and dropping all rotating terms, this leads to the additional term in the Hamiltonian of Eq.~\eqref{eq:hamil}~\cite{PhysRevA.86.013814} 
\begin{align}
H_{K} = \sum_{kl} K_{kl} a_k\dag a_k^{\phantom{\dagger}} a_l\dag a_l^{\phantom{\dagger}}, \label{eq:Hkerr}
\end{align}
where the self- ($K_{kk}$) and cross-Kerr ($K_{kl}$) coefficients can be expressed as
\begin{align}
K_{kl} = -\frac{2-\delta_{kl}}{4L_J \varphi_0^2} \frac{ \tilde{\omega}_k \tilde{L}_k}{2} \frac{ \tilde{\omega}_l \tilde{L}_l}{2} \sum_{n=0}^{N-1} \Delta \phi_k (n)^2 \Delta \phi_l (n)^2, \label{eq:K}
\end{align}
with $\Delta \phi_k(n) \equiv {\phivec}_{k}(n) - {\phivec}_{k}(n+1)$. It follows immediately that $K_{kl} = -2\sqrt{K_{kk} K_{ll}}$~\cite{PhysRevLett.108.240502}. 

We note that the Hamiltonian of Eq.~\eqref{eq:Hhat} was obtained by first finding the qubit-renormalized array modes which were used as a convenient basis. With this approach, the modes already take into account the capacitances $C_s$ and $C_q$ which may be much larger than the array capacitances and hence significantly change the mode structure. With this choice, the mode-mode-couplings , $G_{kl}$, are then very small and can be ignored. Another approach to find the system Hamiltonian would be to first diagonalize the Lagrangian without coupling to $\dot\phi_{qb}$ and then reintroduce this coupling. Such an approach would lead to much larger mode-mode-coupling which would then have to be taken into account by exact diagonalization. Both approaches, in the end, leads to equivalent coupling strengths between the array modes and the qubit, $g_k$.

Before concluding this section, we note that the distinction between the resonator and the transmon in the system Hamiltonian may seem artificial. After all, the split into a transmon degree of freedom and array degrees of freedom is unnecessary to calculate the eigenfrequencies of the combined system. This distinction is, however, useful since one of the modes of the total system, the qubit mode, inherits the most from the transmon's large nonlinearity. This can be made more apparent by replacing the qubit junction by a SQUID. For symmetric junctions, this leads to the replacement $E_J \rightarrow E_{J} \cos(\Phi_x / 2 \varphi_0)$, with $\Phi_x$ the external flux, in the qubit Hamiltonian $H_{qb}$. In this situation, the qubit mode is widely flux tunable while the array modes have, following our treatment, no explicit dependence on flux. This also affects the qubit-array coupling which, in the transmon regime, now takes the form
\begin{align}
g_k (\Phi_x) \approx \cos(\Phi_x / 2 \varphi_0)^{1/4} \; g_k(\Phi_x = 0), \label{eq:couplingtune}
\end{align}
with $ g_k(\Phi_x = 0)$ given by Eq. \eqref{eq:g}. As will be explored below, replacing the qubit junction by a SQUID also provides a tool for initiating dynamics in the system.

\section{Ultrastrong coupling with a transmon}
\label{sec:parameters}

To investigate how strongly the transmon can be coupled to the array, we now focus on the lowest mode of the resonator, $k=0$. Indeed, this mode is expected to have the largest zero-point fluctuations as characterized by $g_0/\omega_0 \propto \sqrt{Z_0}$ (from now on we write $\omega_k$, however, still referring to $\tilde{\omega}_k$ calculated in Sec.~\ref{sec:lagran}). 
To reach a large value of $Z_0$, the array junctions must be of large Josephson inductance $L_J$ and of small capacitance to ground $C_0$. Moreover, since $g_0$ naturally depends on the coupling capacitor, $C_q$, it is also useful to make this capacitance large. However, a change in $C_0$ and $C_q$ does not only change $g_0$, but it also influences other system parameters such as the transmon anharmonicity, $E_C$, the transmon frequency, $\omega_a$, the mode frequencies, $\omega_k$, and the other mode couplings, $g_k$. Our approach to maximize the coupling strength is thus to fix the qubit anharmonicity $E_C$ and the mode frequency $\omega_0$ for fixed values of the capacitance $C_0$. The coupling $g_0$ is then optimized numerically  by varying the rest of the system parameters. As will be clear below, $C_q$ is not part of this optimization but we will varied to find an explicit dependence of $g_0$ on this capacitance. This approach does not guarantee the globally maximal coupling strength, but it is sufficient to identify parameters that yield a transmon in the ultrastrong coupling regime. 

\begin{figure}[t]
\includegraphics[width=\linewidth]{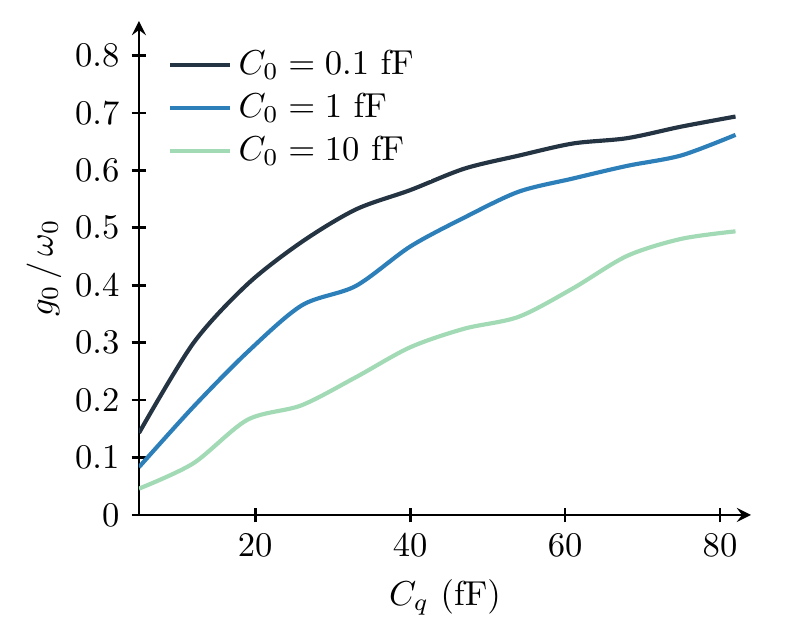}
\caption{Coupling strength, $g_0$, in units of $\omega_0$ as a function of the coupling capacitance $C_q$ as obtained by numerical optimization. From top to bottom the  the parasitic capacitance $C_0$ is increase, with $0.1$ fF at the top followed by $C_0 = 1$ fF and finally $C_0 = 10$ fF. The other parameters are listed in the text.} \label{fig:gc}
\end{figure}

For the numerical examples presented below, we fix the first mode frequency to $\omega_0 = 2\pi \times 2$ GHz, while the transmon anharmonicity is fixed to $E_C = 2\pi \times 300$ MHz with Josephson energy $E_J / E_C = 50$. With a small value of this mode frequency, it is possible to reach a large $g_0/\omega_0$ ratio even with a moderate value of $g_0/2\pi \sim 1$~GHz. In turn, this means that non-adiabatic changes of parameters are possible with realistic flux modulations, allowing for the observation of ultrastrong dynamics. With these choices, Fig.~\ref{fig:gc} shows the results of a numerical optimization of the coupling strength as a function of $C_q$ and for different values of $C_0$. As expected, increasing $C_q$ leads to an increase of $g_0$. The observed oscillations in the coupling strength are due to local maxima in the numerical optimization. The results of Fig.~\ref{fig:gc} highlight that it is possible to reach the ultrastrong coupling regime with a large range of parameters. Finally, we note that the mode frequency $\omega_0$ was chosen to be small, but still large enough to avoid important thermal photon population. 

While generating non-trivial dynamics is the objective here, it also important to have a readout mechanism to probe this dynamic. Because of the photon-number dependent frequency shift resulting from cross-Kerr coupling, it is possible to use a second mode to probe the photon population of the fundamental mode~\cite{PhysRevLett.109.137002}. Therefore, another design objective is to have a large cross-Kerr coupling between modes.

To reach these objectives, we take as parameters:
%\begin{subequations}
\begin{align*}
& N = 145, && L_J = 1.5 \text{ nH}, \\
& C_0 = 0.1 \text{ fF}, && C_q = 85 \text{ fF}, \\
& C_i = 26 \text{ fF}, && C_J = 30  \text{ fF}, \\
& C_e = 72 \text{ fF}, && C_s = 63  \text{ fF},
\end{align*}% \label{eq:params}
%\end{subequations}
which lead to
%\begin{subequations}
\begin{align*}
& \omega_0/2\pi = 2 \text{ GHz}, && g_0/\omega_0 = 0.61, \\
& \omega_1/2\pi = 8.8 \text{ GHz}, && g_1/\omega_1 = 0.11, \\
& \omega_2/2\pi = 14.45 \text{ GHz}, && g_2/\omega_2 = 0.04, \\
& K_{00}/2\pi =  -0.03 \text{ MHz}, && K_{22}/2\pi = -2.46 \text{ MHz}, \\
& K_{02}/2\pi = -0.54 \text{ MHz}, && \wa/2\pi = 5.7 \text{ GHz},
\end{align*}% \label{eq:params2}
%\end{subequations}
with $\wa$ the transmon qubit frequency at $\Phi_x = 0$. Furthermore we take the resonator decay rate $\kappa = {2\pi \times}50$~kHz, corresponding to the losses observed in Ref.~\cite{PhysRevLett.109.137002}. The qubit decay rate and the pure dephasing rate are taken as $\gamma = \gamma_\phi = {2\pi \times}50$~kHz, values that are routinely observed for flux tunable transmons~\cite{mlynek2014observation}. With these choices, the 0th mode is well within the ultrastrong coupling regime while the 1st mode is on the edge of that regime. Moreover, the 2nd mode is both outside the ultrastrong coupling regime and is far-detuned from the qubit. As a result, the dispersive coupling of that mode to the qubit is vanishingly small. On the other hand, as desired the 2nd mode has a significant cross-Kerr coupling to the 0th mode allowing for photon population readout.

\section{Dynamics in the ultrastrong coupling regime}
\label{sec:dynamics}

In this section, we present numerical results of the dynamics for the system with the above parameters and in the presence of damping. Because of the breakdown of the rotating-wave approximate in the ultrastrong coupling regime, it is not possible to use the standard quantum optics master equation~\cite{PhysRevA.84.043832}. We instead use a master equation derived in the instantaneous eigenbasis $\{\ket{j(t)}\}$ of the full system Hamiltonian including Kerr nonlinearity. Following Ref.~\cite{PhysRevA.84.043832}, this master equation reads
\begin{align}
\dot{\rho} &= \frac{-i}{\hbar} [H,\rho] + \sum_{j,k\neq j} \Gamma_{\phi}^{jk} \mathcal{D}\big[ \ket{j}\bra{k} \big]\rho\nonumber \\
&\qquad + \sum_{j, k>j} \big(\Gamma_\kappa^{jk} + \Gamma_\gamma^{jk}\big) \mathcal{D}\big[ \ket{j}\bra{k} \big] \rho   \nonumber \\
&\qquad  + \mathcal{D}\Big[ \sum_j \Phi^j \ket{j}\bra{j} \Big]\rho  , \label{eq:dressedme}
\end{align}
with \mbox{$\mathcal{D}[ O ]\rho = O\rho O\dag - \frac{1}{2}(O\dag O \rho + \rho O\dag O)$}. This equation describes incoherent transitions and dephasing of the system eigenstates with the relaxation rates 
\begin{align}
\Gamma_\gamma^{jk} &= \gamma \,| \bra{j} b\dag + b \ket{k}|^2, \\
\Gamma_\kappa^{jk} &= \kappa \,| \bra{j} a_0\dag + a_0 \ket{k}|^2,
\end{align}
the dephasing rates
\begin{align}
\Phi^j = \sqrt{\frac{\gamma_\phi}{2}} |\bra{j}b\dag b\ket{j}|^2,
\end{align}
and the dephasing-induced relaxation rates
\begin{align}
\Gamma_\phi^{jk} = \frac{\gamma_\phi}{2} | \bra{j} b\dag b \ket{k} |^2.
\end{align}
In the above expressions, $a_0$ ($a_0\dag$) refers to the fundamental mode annihilation (creation) operator and $b$ ($b\dag$) to the qubit lowering (raising) operator. With these forms for the rates, the equilibrium state of Eq.~\eqref{eq:dressedme} is the ground state of the coupled system~\cite{PhysRevA.84.043832}. In contrast, the quantum optics master equation would bring the system to the ground state of the uncoupled system, a state which is far from the true ground state in the ultrastrong coupling regime.

\subsection{Non-adiabatic generation of photons}

\begin{figure}[t]
\includegraphics[width=\linewidth]{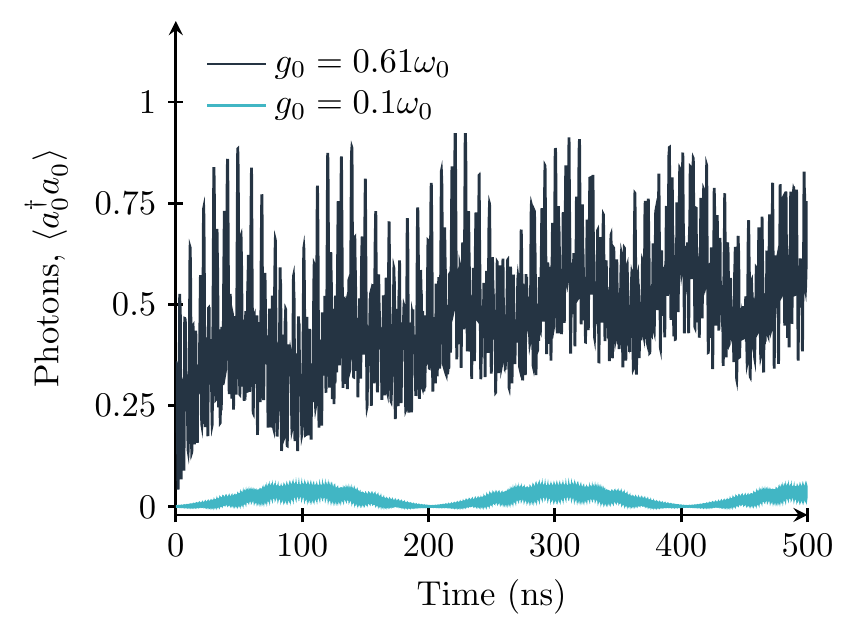}
\caption{Average photons number in the fundamental mode of the array as a function of time. The coupling strength is $g_0(\Phi_x = 0) =0.61\omega_0$ (dark blue line) and $0.1\omega_0$ (light blue line). The external flux is $\Phi_x = 0.35 \Phi_0 \times \cos(\omega_d t)$ with $\omega_d = 2\pi \times 1.5$ GHz. The other parameters are listed in the text.} \label{fig:omega15}
\end{figure}

As already mentioned, an important feature of the Jaynes-Cummings Hamiltonian is that its ground state is that of the uncoupled system. As a result, the nature of this ground state does not change with system parameters. In other words, if prepared in its ground state, a system described by the Jaynes-Cummings Hamiltonian will remain in the vacuum state under parametric modulations. 

In contrast, the ground state, $\ket{j=0}$, of the Rabi Hamiltonian can be approximated as~\cite{PhysRevA.84.043832}
\begin{equation}
\ket{j=0} \approx \left(1-\frac{\lambda^2}{2}\right) \ket{00} - \Lambda \ket{11}+ \xi \sqrt 2 \ket{02} \label{eq:rabi_gs}
\end{equation}
to second order in $\Lambda = g/(\wa+\wc)$ and with $\xi = g\Lambda/2\wc$. On the right-hand-side of this expression, the first index in the states refers to the qubit and the second to the photon number. Equation \eqref{eq:rabi_gs} makes it clear that the ground state of $H_\mathrm{Rabi}$ depends on the system parameters and, moreover, has a finite average photon number. Since the master equation Eq.~\eqref{eq:dressedme} relaxes the system back to $\ket{j=0}$, these photons do not decay out of the cavity and are consequently difficult to observe.

Here we propose to take advantage of the dependence of $\ket{j=0}$ on the system parameters to observe a signature of these photons. Indeed, a non-adiabatic change of the system parameters should lead to a change of the average photon population under $H_\mathrm{Rabi}$ while it should have no effect under $H_\mathrm{JC}$. As alluded to earlier, this photon population can then be probed by taking advantage of the cross-Kerr coupling between array modes. For the photon population to change under parametric modulations, this modulation must, however, be non-adiabatic. This is possible in this system and with the parameters of Sec.~\ref{sec:parameters} because of the small mode frequency $\omega_0$ and therefore the reasonably small $g_0$ required to reach ultrastrong coupling.

To realize this, we modulate the flux through the transmon's SQUID loop as 
\begin{align}
\Phi_x(t) = \Phi_x^a \cos(\omega_d t)
\end{align}
to induce non-adiabatic dynamics~\cite{huang2016manipulating}. To reach measurable photon populations, large flux modulations $\gtrsim 0.1 \Phi_0$ are required. While this modulation amplitude is larger than what is typically used in flux-pumped Josephson parametric amplifiers, similar amplitudes have already been demonstrated experimentally~\cite{wilson2011observation}.

\begin{figure}[b]
\includegraphics[width=\linewidth]{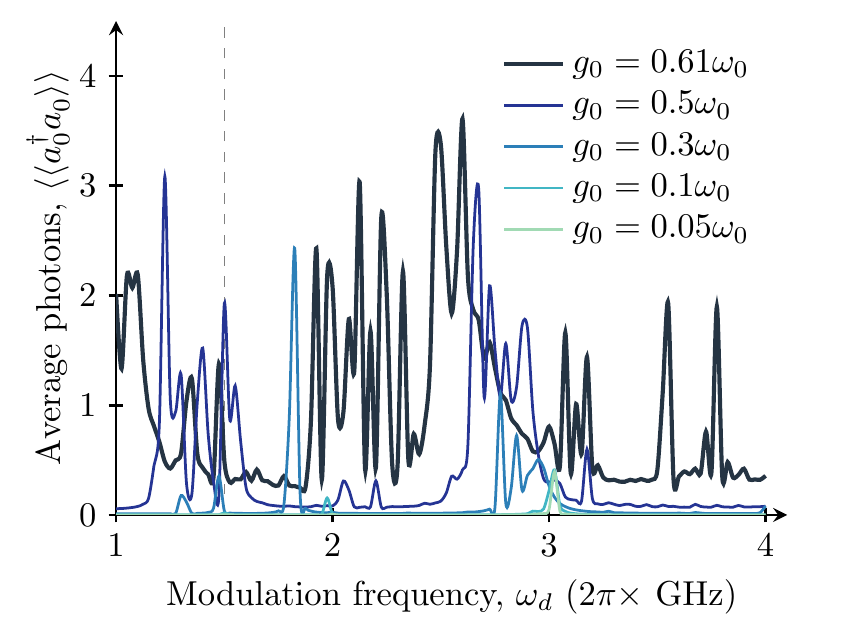}
\caption{Time averaged average photon number for different modulation frequencies and coupling strengths. The lines from light to dark indicates stronger coupling, $g_0$, as indicated by the legend. The vertical dashed line indicates the value used in Fig.~\ref{fig:omega15}.
} \label{fig:avpho}
\end{figure}

Because of the change in system parameters under this flux modulation, the overlap between the instantaneous ground state at a given time, $\ket{j=0(t)}$, and the $j'$th excited state at a later time $t'$, $\ket{j'(t')}$ will in general be non-zero,
\begin{align}
\braket{j=0(t)}{j'(t')} \neq 0,
\end{align}
a result that holds only when the RWA is not valid. This implies that flux modulations can excite the system away from the ground state. An example of this non-adiabatic dynamic is presented in Fig.~\ref{fig:omega15} which shows the photon population as a function of time as obtained by numerical integration of Eq.~\eqref{eq:dressedme} with the parameters of Sec.~\ref{sec:parameters} and a modulation frequency of $\omega_d = 2\pi \times 1.5$~GHz. In these simulations, the system was first initializing in the ground state $\ket{j=0}$. Importantly, the drive frequency does not correspond to a resonance frequency of the coupled system and is therefore not expected to directly drive specific system transitions. Despite this, a consequential photon population is observed for $g_0/\omega_0 = 0.61$ (dark-blue line). On the other hand, a weaker coupling  of $g_0/\omega_0 = 0.1$ (light blue line) for which the Jaynes-Cummings Hamiltonian is expected to be a good approximation shows a much smaller average photon population.

To further illustrate this point, Fig.~\ref{fig:avpho} shows the time-averaged photon number as a function of the modulation frequency $\omega_d$ and for different drive strengths. Again, we observe that, for the small coupling strengths, very few photons are generated and this occurs only at well-defined resonances. In the ultrastrong coupling regime photons are, however, generated for a large range of frequencies. For the strongest coupling of $g_0/\omega_0 = 0.61$ (dark blue line), photons are observed for all drive frequencies. This confirms that photons are not generated by directly exciting a transition of the static system, but are due to the non-adiabatic change of the ground state. An analogy can be drawn to multi-passage Landau-Zener transitions~\cite{shevchenko2010landau}. These transitions appear when the parameters of a two-level system is changed in a non-adiabatic fashion through an avoided crossing. A similar effect is observed here with a non-adiabatic change in the ground state. The transmon-array system has, however, a complex level structure where the Landau-Zener results cannot be explicitly applied.

\subsection{Photon population measurement}
\label{sec:readout}

To measure the photon population in the ultrastrongly coupled mode $k=0$, we take advantage of the cross-Kerr coupling between modes $k=0$ and 2. This coupling was already used experimentally to characterize a junction array~\cite{PhysRevLett.109.137002}. Ignoring the other array modes, this coupling takes the form
\begin{align}
H_K = K_{02} a_0\dag a_0 \, a_2\dag a_2, \label{eq:crossK}
\end{align}
with $K_{02} = -4\sqrt{K_{00} K_{22}}$. Photon population in mode 0 will shift the second mode frequency by $K_{02} a_0\dag a_0$, a shift that can be resolved by probing mode 2.

\begin{figure}[t]
\includegraphics[width=\linewidth]{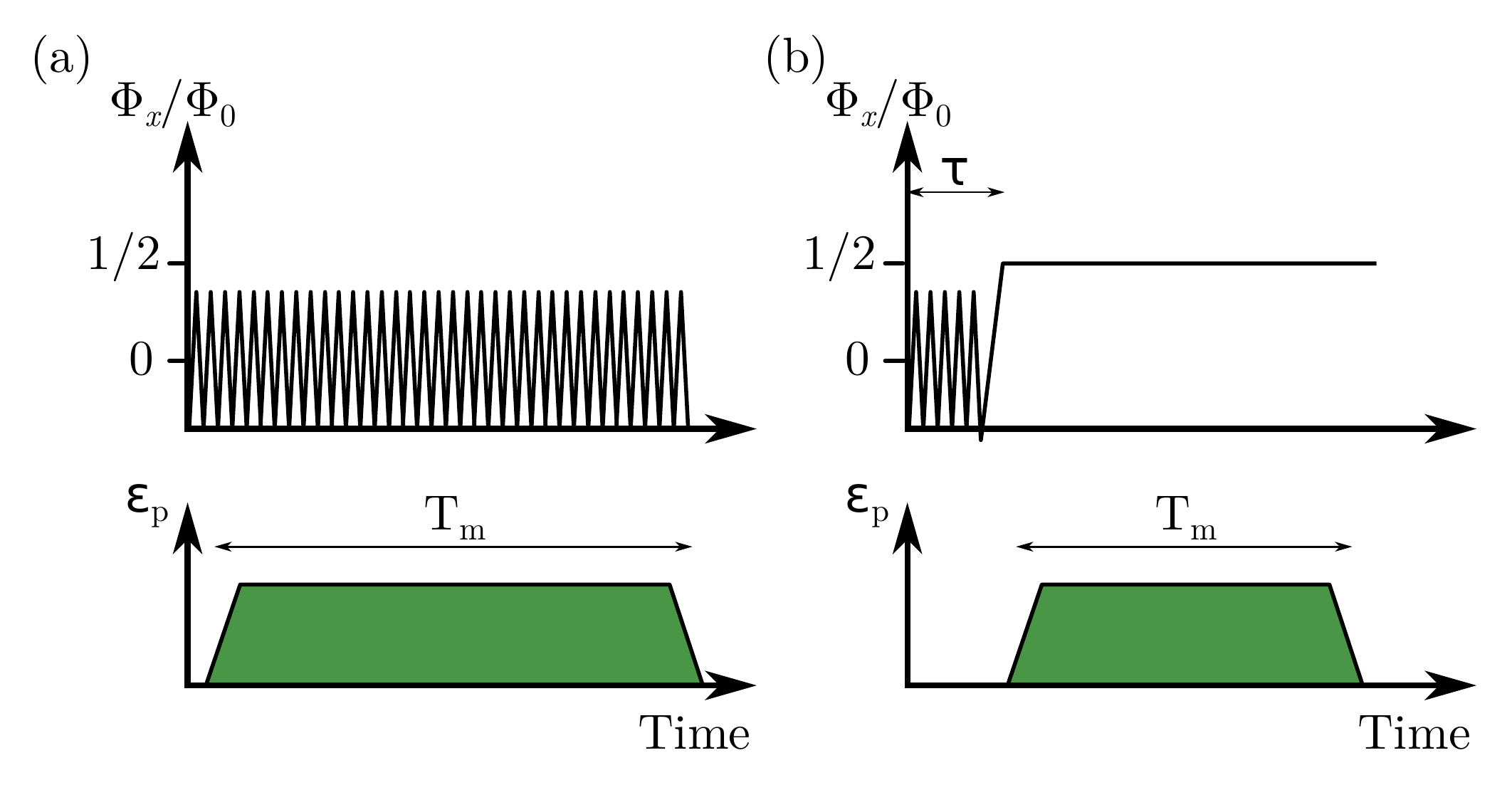}
\caption{Cross-Kerr readout schemes to probe the photons generated by the ultrastrong coupling dynamics. In (a) we apply the modulation of the transmon continuously while probing a higher mode of the Kerr resonator. In (b) we only modulate for a time $\tau$ followed by a probing of a higher mode.} \label{fig:readout}
\end{figure}

The general approach is now to apply a coherent drive, $H_p = \varepsilon_p (a_2 + a_2\dag)$, on resonance with the probe mode via the input port $C_i$ (see Fig.~\ref{fig:lumped}). The signal reflected from this port is then continuously monitored. Similarly to dispersive qubit readout~\cite{PhysRevA.69.062320},  the photon number, $\exv{a_0\dag a_0}$, can be determined by homodyne measurement of the field amplitude $a_2$. The integrated homodyne signal can be expressed as 
\begin{align}
M_{\varepsilon_p} = \sqrt{\kappa_2} \int_\tau^{\tau+T_m} [a_\mathrm{out}(t) + a_\mathrm{out}\dag(t)] \, dt
\end{align}
with $T_m$ the integration time and $\tau$ the initial time of the integration. In this expression, $a_\mathrm{out}(t) = \sqrt{\kappa_2}a_2(t) + a_\mathrm{in}(t)$ is the output field~\cite{gardiner2004quantum}, with $a_\mathrm{in}$ the input noise of the vacuum respecting $[a_\mathrm{in}(t),\,a_\mathrm{in}\dag(t')] = \delta(t'-t)$, and $\kappa_2$ the decay rate of mode $k=2$. The ability for such a measurement to distinguish the state from the system with no flux modulation, ie. no ultrastrong dynamics, is captured by the signal-to-noise ratio (SNR). Following Ref.~\cite{RevModPhys.82.1155, PhysRevLett.115.093604}, the SNR can be expressed as
\begin{align}
\text{SNR}_{T_m} = \frac{|\exv{M_{\varepsilon_p}} - \exv{M_0}|}{\sqrt{\exv{\tilde{M}_{\varepsilon_p}^2} + \exv{\tilde{M}_0^2}}}, \label{eq:snr}
\end{align}
with $\tilde{M}_{\varepsilon_p} = M_{\varepsilon_p} - \exv{M_{\varepsilon_p}}$ and $M_{0}$ corresponds to the same measurement without flux-modulation, $\Phi_x = 0$. 

As illustrated in Fig.~\ref{fig:readout}, two approaches are considered. In the first approach, depicted in panel (a), the probe field is monitored while continuously modulating the qubit flux. For simplicity, the nonlinearity of the probe mode is ignored and the photon number $a_0\dag a_0$ is taken to be a classical number. Then, the equation of motion for $a_2$ reads
\begin{align}
\dot{a}_2 = -iK_{02} \exv{a_0\dag a_0} a_2 - \frac{\kappa_2}{2} a_2 - i \varepsilon_p + \sqrt{\kappa_2} a_\mathrm{in}
\end{align}
which as the steady state solution
\begin{align}
a_2^\mathrm{s} = \frac{\sqrt{\kappa_2} a_\mathrm{in} - i\varepsilon_p}{iK_{02}\exv{a_0\dag a_0}  + \kappa_2/2}. \label{eq:a2ss}
\end{align}
To obtain a simple estimate for the SNR, we use the values of $\exv{a_0\dag a_0}$ oscillating between 0.2 and 0.8 shown in Fig.~\ref{fig:omega15} and integrate the signal taking $\tau=100$ ns to go beyond the initial ring up dynamics. For the parameters presented in Sec.~\ref{sec:parameters}, together with \mbox{$\varepsilon_p = 2\pi \times 2$~MHz} and $\kappa_2 = 2\pi \times 0.35$~MHz, this yields a SNR larger than 1 for an integration time $T_m \approx \kappa_2^{-1}$. A larger SNR can be obtained by longer integration times, however, the ultrastrong dynamics will eventually dephase due the dephasing rates $\Phi^j$. Using Eq.~\eqref{eq:a2ss} the value of $\exv{a_0\dag a_0}$ is estimated and we recover, as desired, the numerical time-averaged results shown in Fig.~\ref{fig:avpho}. In this analysis we neglected the self-Kerr nonlinearity, $K_{22} = 2\pi\times -2.4$ MHz, but in general similar results for the cross-Kerr probing can be obtained by including the nonlinearity in the analysis~\cite{andersenkamal}.

An alternative method to map the dynamics shown in Fig.~\ref{fig:omega15} is sketched in Fig.~\ref{fig:readout}(b). In this approach, the qubit flux is modulated for a time $\tau$ around $\Phi_x = 0$ with an amplitude of $\Phi_x = 0.35\Phi_0$. After this initial period of ultrastrong dynamics, the flux is rapidly increased to $\Phi_x = \Phi_0/2$ in a time span of one full period of oscillation, $2\pi/\omega_d \approx 0.66$ ns.  At that point, the qubit has a vanishingly small transition frequency and is uncoupled from the array, see Eq.~\eqref{eq:couplingtune}. Now, with the coupling to the qubit absent, the population of mode $a_0$ simply decays to the vacuum state at a rate $\kappa_0$. 
Again, it is worth emphasizing that due to the choice of a small resonator frequency and, thus, low coupling, this change in flux is fast enough to maintain the photon number in the $a_0$ mode. Using the same parameters as in Fig.~\ref{fig:omega15}, we numerically integrate Eq. \eqref{eq:snr} and find, taking into account array damping, a maximal \mbox{SNR$_{T_m} \approx 0.5$} for a measurement time \mbox{$T_m \approx 6$ $\mu$s}. For larger measurement times, the signal will be dominated by noise because the photon population of the $a_0$ mode have decayed. This estimate is obtained from numerical integration including cross-Kerr coupling given by Eq.~\eqref{eq:crossK} and self-Kerr nonlinearities for both modes. As above, from the measured signal, the detuning of the probe mode $a_2$ from its bare frequency $\omega_2$ can now be inferred. Using the inferred probe detuning, the photon population in mode $k=0$ can then be estimated. Therefore, by treating the average photon number as an unknown parameter, standard parameter estimation techniques \cite{wiseman2009quantum, PhysRevA.87.032115} can be used and the dynamics generated during the ultrastrong coupling is observed. We expect that a measurement of the dynamics can be obtained by only a few experimental runs for each value of~$\tau$.

\section{Conclusion}
\label{sec:conc}

In conclusion, we have shown that it is possible to reach the ultrastrong coupling regime of light-matter interaction by coupling a transmon qubit to a high impedance mode realized by an array of weakly nonlinear Josephson junctions. Using realistic system parameters, we find coupling strengths as large as 0.6 times the system frequency. By working with a low frequency mode of the array, this ultrastrong coupling is obtained for moderate values of the coupling. This is an important advantage of our proposal. Indeed, with this choice, we have shown that realistic modulations of the qubit parameters are sufficient to results in dynamics of the system that is distinctive of the ultrastrong coupling regime. Moreover, we have shown how this dynamic and the corresponding photon population can be probed by taking advantage of the multi-mode structure of the array. These results show the possibility to probe the complex dynamics of the ultrastrong coupling regime, opening a new window on this unconventional regime of quantum optics.

\section*{Acknowledgements}
The authors are grateful to J.~Bourassa, N.~Didier, A.~Kamal and N.~Roch for discussions. Financial support from the Villum Foundation Center of Excellence, QUSCOPE, from the Danish Ministry of Higher Education and Science and by NSERC is acknowledged. This research was undertaken thanks in part to funding from the Canada First Research Excellence Fund.

\bibliography{bt_ultra}

\end{document}